\documentclass[preprint,12pt]{elsarticle}

\usepackage{amssymb}

\journal
{New astronomy}

\begin{document}

\begin{frontmatter}



\title{Signal of quark deconfinement in thermal evolution neutron stars with deconfinement heating}


\author{Miao Kang , Xiao-Dong Wang}

\address{The College of Physics and Electronics, Henan
University, Kaifeng, 475004,P.R.China}
\author{Xiao-Ping Zheng}
\address{The Institute of Astrophysics, Huazhong Normal University,
Wuhan 430079, P.R.China}
\begin{abstract}
 As neutron stars spin-down and contract, the
deconfinement phase transition can continue to occur, resulting in
energy release(so-called deconfinement heating) in case of the
first-order phase transition. The thermal evolution of neutron
stars is investigated to combine phase transition and the related
energy release self-consistently. We find that the appearance of
deconfinement heating during spin-down result in not only the
cooling delay but also the increase of surface temperature of
stars. For stars characterized by intermediate and weak magnetic
field strength, a period of increasing surface temperature could
exist. Especially, a sharp jump in surface temperature can be
produced as soon as quark matter appears in the core of stars with
a weak magnetic field. We think that this may serve as evidence
for the existence of deconfinement quark matter. The results show
that deconfinement heating facilitates the emergence of such
characteristic signature during the thermal evolution process of
neutron stars.

\end{abstract}


\end{frontmatter}


\section{Introduction}
Fundamental properties of supranuclear matter in the cores of
neutron stars, such as the chemical composition and the equation
of state, are still poorly known. Simulations of the thermal
evolution of neutron stars confronted with soft X-ray, extreme UV,
and optical observations of thermal photon flux emitted from their
surface provide the most valuable information about the dense
matter in the interior of these stars.



As the interior density gradually increases of neutron stars,
deconfinement phase transition continuously takes place inducing
not only structural changes but also energy release in case of a
first-order phase transition. The generation of energy increases
the internal energy of the star which is called deconfinement
heating (Haensel $\&$ Zdunik \cite{91hae},Yu \& Zheng
\cite{06yu},Kang \& Zheng \cite{07kan}). The thermal evolution of
neutron stars is connected with their spin-down and the resulting
changes in structure and chemical composition (from nucleon matter
to deconfined quark matter)have been investigated in our work. We
have investigated the thermal evolution of neutron stars with such
a deconfinement phase transition(Kang \& Zheng \cite{07kan}). The
results show that deconfinement heating delays dramatically the
cooling of neutron stars, which have a higher surface temperature
compared with traditional cooling for the same age stars.


Many efforts are devoted to explore the observational signal of a
deconfinement phase transition which have been suggested
  in the form of characteristic changes of observables, such as the pulse timing(Glendenning et al.\cite{97gle2},Chubarian et al.\cite{00chu},Poghosyan et al.\cite{01pog}),
   brightness (Dar \&DeR\'{u}jula \cite{00dar})and surface temperature(Schaab et al.\cite{96sch}, Blaschke et al.\cite{00bla},Yuan \& Zhang \cite{99yua},
   Stejner et al.\cite{08ste}) of the pulsars during their
   evolution. The surface temperature changes when the quark matter appears in the cores of neutron
   stars. We explore the deconfinement signature by studying the changes of
surface temperature in the thermal evolution process of neutron
stars.

In this paper, we reinvestigate the thermal evolution of neutron
stars and look for a characteristic change in the surface
temperature with deconfinement heating. The released energy can
also be estimated as a function of the change rate of the
deconfinement baryon number using the parameterized approach. The
neutron stars containing quark matter are called hybrid stars. We
take Glendenning's hybrid stars model (Glendenning \cite{97gle})
based on the perturbation theory developed by Hartle\cite{67har}
to study the rotational evolution structure of stars. The Argonne
$V18+\delta\upsilon+UIX^{*}$ model(APR)(Akmal et al. \cite{98akm})
of hadronic matter and the MIT bag model of quark matter are used
to construct the model of stars, but the medium effect of quark
matter has been considered in quasi-particle description(Schertle
et al. \cite{97sch}).

\section*{2. Deconfinement phase transition and neutron stars structure}

Early works on deconfinement phase transition, the Maxwell
construction, show a sharp transition taking place between the two
charge-neutral hadron and the quark phases(Baym \& Chin
\cite{76bay}).
 In the 1990s, Glendenning(Glendenning \cite{92gle}, \cite{97gle}) pointed
out that this assumption was too restrictive. More generally, the
transition can occur through the formation of a mixed phase of
hadron matter and quark matter, with the total charge neutrality
being achieved by a positively charged amount of hadronic matter
and a negatively charged amount of quark matter. Following
Glendenning's model, we use a standard two-phase description of
the equation of state(EOS) through which the hadron and quark
phases are modelled separately.  The resulting EOS of the mixed
phase is obtained by imposing Gibbs's conditions for phase
equilibrium with the constraint that the baryon number and the
electric charge of the system are conserved to the neutron star
matter.

The Gibbs condition for chemical and  mechanical equilibrium at
zero temperature between the two phases reads
\begin{equation}
p_{HP}(\mu_{n},\mu_{e})=p_{QP}(\mu_{n},\mu_{e})
\end{equation}
where $p_{HP}$ is the pressure of confined hadron phase and
$p_{QP}$ is the pressure of deconfined quark phase.

The conservation laws can be imposed by introducing the quark
fraction $\chi$ defined as $\chi=V_{Q}/V$. Only two independent
chemical potentials remain according to the corresponding two
conserved charges of the $\beta$-equilibrium system. The total
baryon number density $\rho_{B}$ is
\begin{equation}
\rho_{B}=\frac{N_{B}}{V}=\chi \rho_{QP}+(1-\chi) \rho_{HP}
\end{equation}
the total electrical charge is
\begin{equation}
0=\frac{Q}{V}=\chi q_{QP}+(1-\chi) q_{HP}
\end{equation}
and the total energy density is
\begin{equation}
\epsilon=\frac{E}{V}=\chi \epsilon_{QP}+(1-\chi) \epsilon_{HP}
\end{equation}
 Using the Eqs.(1), (2), (3)and (4), we can obtain the EOS of mixed phase matter.
In describing the hadronic part of the neutron star, we adopt the
APR model (Akmal et al. \cite{98akm}). For the EOS, it is based on
the models for the nucleon interaction with the inclusion of a
parameterized three-body force and relativistic boost corrections.
We use the EOS of an effective mass bag-model for the quark matter
part of the neutron star(Schertle et al. \cite{97sch}).

In Fig.1, we show the model EOS with deconfinement transition,
which is the typical scheme of a first-order transition. The phase
transition construction in a two-component system leads to a
continuous increase in energy per baryon in the mixed phase with
increasing density. It is well known that hadron matter is the
most stable phase at lower densities, and that quark matter is the
most favorite phase at higher densities. Meanwhile, the mixed
phase has the lowest energy at intermediate densities. We choose
the parameters for quark matter EOS with s quark mass
$m_{s}=150MeV$, coupling constant $g=3$ and different bag constant
$B=85MeVfm^{-3}$, $B=108MeVfm^{-3}$, $B=136MeVfm^{-3}$
respectively.

With the EOS presented above, we are ready to study the structural
evolution of the rotating neutron stars.
 In this paper, we apply Hartle's approach (Hartle \cite{67har}) as in Kang \& Zheng \cite{07kan} to investigate
the structure of the stars. By treating a rotating star as a
perturbation on a non-rotating star and by expanding the metric of
an axially symmetric rotating star in even powers of the angular
velocity $\Omega$, we can obtain the structure of the rotating
stars.

The resulting gravitational masses of neutron stars for
$B=108MeVfm^{-3}$ are shown in Fig.2, as a function of central
baryon density for static stars as well as for stars rotating with
the maximum rotation frequency $\nu_{k}$. The solid almost
horizontal connect configurations with the same total baryon
number. In order to explore the increase in central density due to
spin down, we created sequences of neutron star models. A Model in
particular sequence has the same constant baryon number,
increasing central density and decreasing angular velocity.

\section*{3. Deconfinement heating}
There is deconfinement heating production due to spin-down in
neutron stars due to the nuclear matter continuously converting
into quark matter. The released energy had been estimated as a
function the rate of change the deconfinement baryon number using
the parameterized approach(Kang \& Zheng \cite{07kan}). Recently
we studied the mechanism of energy release in detail(Kang et al.
\cite{07mia}). Through studying a random process of infinitesimal
compression for the mixed phase region, we can calculate the
energy release per baryon using the following formula
\begin{equation}
\delta \tilde{e}-\delta e=(\frac{\delta
\tilde{e}}{\delta\rho_{B}}-\frac{\delta e}{\delta\rho_{B}})\delta
\rho_{B},
\end{equation}
where $\frac{\delta\tilde{e}}{\delta\rho_{B}}$ denotes the
enthalpy change per baryon .

The deconfinement heating is coupled with the rotation evolution
of neutron stars. Combining the energy change with the
evolutionary structure of neutron stars, we get the total heat
luminosity(Kang et al. \cite{07mia}\cite{08kan})
\begin{equation} H_{dec}=\int
\frac{de}{dv}\dot{v}(t)\rho_{B}dV
\end{equation}
where $v$ is the rotation frequency of the star. The spin-down of
stars is due to the magnetic dipole radiation. The rotation
frequency is given by
\begin{equation}
\dot{v}=-\frac{16\pi^{2}}{3Ic^{3}}\mu^{2}v^{3}\sin^{2}\theta
\end{equation}
 where $I$ is the stellar
moment of inertia, $\mu=\frac{1}{2}BR^{3}$ is the magnetic dipole
moment, and $\theta$ is the inclination angle between the magnetic
and rotational axes. We now combine the energy release behavior
with spin-down. Our recent work shows that exact calculations
agree well with the earlier order of magnitude estimates and
supports the previous parameterized approach. Because the change
of rotating stars is sufficiently slow, the formula of heat
luminosity can be replaced by the simple parameterized form(Kang
\& Zheng \cite{07kan}),
\begin{equation}
H_{dec}=\bar{q}{{\rm d}N_{Q}\over {\rm d}v}\dot{v}(t)
\end{equation}
where $\bar{q}$, about 0.1MeV, is the mean value of energy release
in the mixed phase and $N_{Q}$ represents the baryon number of
quarks in the interior of the star.

\section*{4.Signal of quark deconfinement and thermal evolution of neutron stars}

The cooling of neutron stars could take place via two channels -
neutrino emission from the entire star and thermal emission of
photons through the transport of heat from the internal layers to
the surface. Neutrino emission is generated in numerous reactions
in the interior of neutron stars, e.g. as reviewed by Page et
al.\cite{05pag}.
 For the calculation of cooling of the hadronic part of the neutron star, we use the main processes
 including the nucleon direct Urca (NDU) and the nucleon modified Urca(NMU) and the nucleon bremsstrahlung(NB).
For the quark matter, we consider the main process: the quark
direct Urca (QDU) processes on unpaired quarks, the quark modified
Urca (QMU) and the quark bremsstrahlung(QB). For pure neutron
stars, the direct Urca reaction (the most efficient) is allowed
only at very high densities because it is impossible to satisfy
the conservation of momentum unless the proton fraction exceed the
value where both the charge neutrality and the triangle inequality
 can be observed (Lattimer et al.\cite{91lat}). However, for neutron stars that contains quark matter,
this is not so because charge neutrality does not have
  to be conserved locally for a mixed phase. Hence the NDU process is active in the mixed phase.

The calculation of the evolution of the thermal energy of neutron
stars (heating and cooling) is achieved by coupling their rotating
structure and deconfinement phase transition. Fig.3 displays the
central density of different masses neutron stars for
$B=108MeVfm^{-3}$, as a function of their rotational frequency. In
Fig.3, the dotted horizontal line indicates the deconfined quark
matter produced and dashed horizontal line indicate the NDU
processes triggered. In the interior of these stars, the emerging
of deconfined quark matter accompanies the gradual energy release
leads to rising of surface temperature, and appearance of the NDU
process can result in a rapid decrease of the temperature during
the spin-down. For $M=1.5,1.55, 1.6 M_{\odot}$ neutron stars, the
deconfinement phase transition occurs during the thermal evolution
process which may lead to appearance of a characteristic
signature. We will now discuss in detail the quark deconfinement
signal.

 We combine the equation of thermal balance with the rotating structure equations of the
stars(Kang \& Zheng \cite{07kan}, Hartle \cite{67har}) and rewrite
the energy equation in the approximation of an isothermal
interior(Glen \& Sutherland \cite{80gle})
\begin{equation}
C_{V}(T_{i},v)\frac{dT_{i}}{dt}=-L_{\nu}^{\infty}(T_{i},v)-L_{\gamma}^{\infty}(T_{s},v)
\end{equation}
\begin{equation}
C_{V}(T_{i},v)=\int_{0}^{R(v)}c(r,T)(1-\frac{2M(r)}{r})^{-1/2}4\pi
r^{2}dr
\end{equation}
\begin{equation}
 L_{\nu}^{\infty}(T_{i},v)=\int_{0}^{R(v)}\varepsilon(r,T)
(1-\frac{2M(r)}{r})^{-1/2}e^{2\Phi}4\pi r^{2}dr
\end{equation}
Where $T_{s}$ is the effective surface temperature,
$T_{i}(t)=T(r,t)$ is the redshifted internal temperature; $T(r,t)$
is the local internal temperature of matter, and $\Phi(r)$ is the
metric function(describing gravitational redshift)(Yakovlev \&
Haensel \cite{03yak}). Furthermore, $L_{\nu}^{\infty}(T_{i},v)$and
$C_{V}(T_{i},v)$ are the total redshifted neutrino luminosity and
the total stellar heat capacity, respectively, which are functions
of rotation frequency and temperature; $c(r,T)$ is the heat
capacity per unit volume. $L_{\gamma}^{\infty}=4\pi R^{2}(v)\sigma
T_{s}^{4}(1-R_{g}/R)$ is the surface photon luminosity as detected
by a distant observer($R_{g}$ is the stellar gravitational
radius). The effective surface temperature which is detected by a
distant observer is $T_{s}^{\infty}=T_{s}\sqrt{1-R_{g}/R}$.
$T_{s}$ is obtained from the internal temperature by assuming an
envelope model(Gudmundsson et al. \cite{83gud}, Potekhin et al
\cite{97pot}). Using to Eqs.(7)-(11), we can simulate the thermal
evolution of neutron stars with deconfinement heating.

  In Fig.4 and Fig.5, we present thermal evolution behavior of a 1.6$M_{\odot}$ neutron star
     for different magnetic fields ($10^{9}-10^{12}$G).
Due to the coupling of thermal evolution and spin-down, all
curves(with deconfinement heating and without deconfinement
heating) show clear magnetic field dependence. During the star's
spin down, deconfined quark matter appears in the core of the star
at a spin frequency of $v=1123$Hz, the surface temperature drops
rapidly when the NDU process(enhanced cooling)
  occurs at a spin frequency of $v=492$Hz.
 It is evident
     that the temperature of the curves with deconfinement heating(solid curves)
    are higher than for the standard cooling scenario without deconfinement(dotted curves).
    We can observe a competition between cooling
and heating processes from the heating curves, where deconfinement
heating can produce a characteristic
     rise of surface temperature and even dominate the history of thermal evolution. Eventually, they reach a thermal equilibrium,
     where the heat generated is radiated away at the same rate from the star surface. We find the weaker magnetic fields have the larger change of
      temperature. The low magnetic field ($10^{9}$G) produces a sharp jump in surface temperature as soon as the deconfinement quark matter
      appearing during spin-down. Intermediate magnetic field ($10^{10},10^{11}$G) lead to slight changes in the temperature, but high magnetic field form
  only the temperature plateau at a time.

In Fig.6, we present the cooling behavior of different masses
stars for magnetic field B=$10^{12},10^{11}$G (left panel) and
magnetic field B=$10^{9},10^{8}$G(right panel) with deconfinement
heating. The observational data, taken from tables 1 and 2 in Page
et al.\cite{04pag}, have been shown in left panel. Comparing with
 previous investigation(Kang \& Zheng \cite{07kan}), we find the
thermal evolution curves of our present work are more compatible
with the observational data(left panel). In our previous work it
seemed that, the NDU processes should be triggered easily in the
model of relativistic mean field (the critical mass for fast
cooling occuring is being low )(Glendenning \cite{97gle}). In our
present study, using the APR EOS, NDU processes can not be
triggered easily in stars which lead to the higher temperatures of
the evolution curves than in the previous cases. For example, the
NDU reactions only appear above 1.56 $M_{\odot}$ in present model.
In the cases of weak magnetic field, stars have high
temperatures($>10^{5}$K) at older ages ($>10^{9}$yrs). We thus
think that
 high temperature of some millisecond pulsars with low magnetic fields (Kargaltsev et al \cite{04kar}), especially for PSR J0437-4715,
  can be explained using the deconfinement heating model of neutron stars.  We can observe that 1.5 $M_{\odot}$
 neutron stars follow a similar thermal evolution track as 1.6 $M_{\odot}$, but there is not a period increasing in temperature
   for 1.7$M_{\odot}$. Through comparing Fig.6 with Fig.3, we find the quark matter to appear at the birth of star for 1.7$M_{\odot}$. For
1.5 $M_{\odot}$ and 1.6 $M_{\odot}$ stars, quark deconfinement
occurs when the central density gradually increases during
spin-down, which results in the temperatures of the stars to
increase rapidly. This is a characteristic signal as quark matter
 arises during the rotational spin-down of stars for weak magnetic case.

\section*{5. Conclusions and discussions}
The chief aim of the present work is to explore the signal of
quark matter appearing
 through theoretical simulation of the thermal evolution curves of neutron stars with deconfinement heating. We have constructed  models of rotating
 neutron stars that follow the mixed phase investigation of Glendenning based on Hartle's perturbative approach.
 The total thermal luminosities have been obtained using the
parameterized approach.

Recently, Stejner et al \cite{08ste} have investigated the
signature of deconfinement with spin-down compression in cooling
neutron stars. A period of increasing surface temperature can be
produced with the introduction of a pure quark core for strongly
superfluid stars of strong and intermediate magnetic field
strength and the latent heat of deconfinement  reinforces the
signature only but is itself relatively less significant. Contrary
to their studies, our results show that deconfinement heating can
drastically affect the thermal evolution of neutron stars. The
rise of surface temperature of cooling stars, as a signature of
quark deconfinement, is derived from the deconfinement heating. It
is noteworthy that a significant rise of the temperature
accompanies the appearance of quark matter at older ages for low
magnetic field stars. This may be a evidence for existence of
quark matter, if a period of rapid heating is observed for a very
old pulsar. Deconfinement heating provides a new way to study the
signal of deconfinement.


We found that the deconfined signal appears for neutron stars of
mass $1.4M_{\odot}\lesssim M\lesssim1.64M_{\odot}$. The influence
of different EOS of the hadron phase and the model parameters of
the quark phase (bag constant B, coupling constant g) on the phase
transition densities, rotational structure of neutron stars and
corresponding internal structure etc have been studied by many
investigators(Schertler et al.\cite{00sch}; Pan et
al.\cite{06pan}). The deconfinement heating rate and mass range of
a deconfinement sinal emerging can be changed with varying of
these parameters. In future, we will systematically investigate
the effect of these parameters which which will be the subject of
our future investigations.



\section*{Acknowledgments}
 This work is supported by NFSC under Grant Nos.10747126 and 10773004.


\begin{figure}
   \centering
   \includegraphics[width=0.7\textwidth]{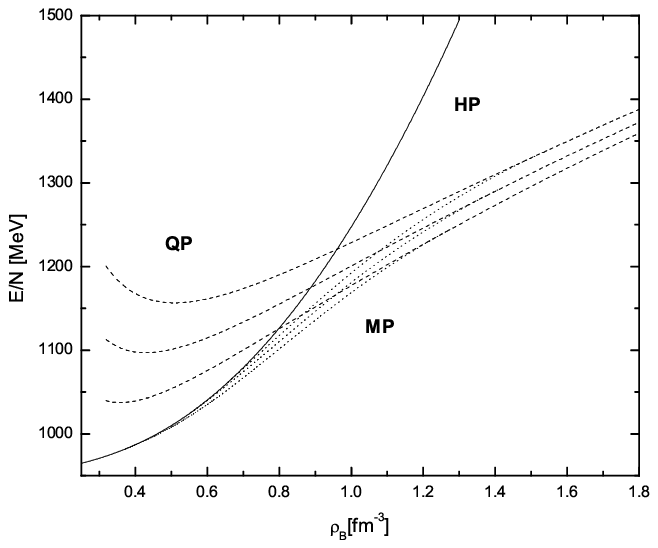}
   \caption{For beta-stable matter, the energy per baryon for the Argonne $V18+\delta\upsilon+UIX^{*}$ model, and quark effective mass MIT bag model
   are shown by full and dashed curves(from top to bottom bag constant B=136,108 and 85 $MeV fm^{-3}$ respectively); the dotted lines correspond to neutral mixtures of
   charged hadron and quark matter}
   \label{Fig:f1}
   \end{figure}

\begin{figure}
   \centering
   \includegraphics[width=0.7\textwidth]{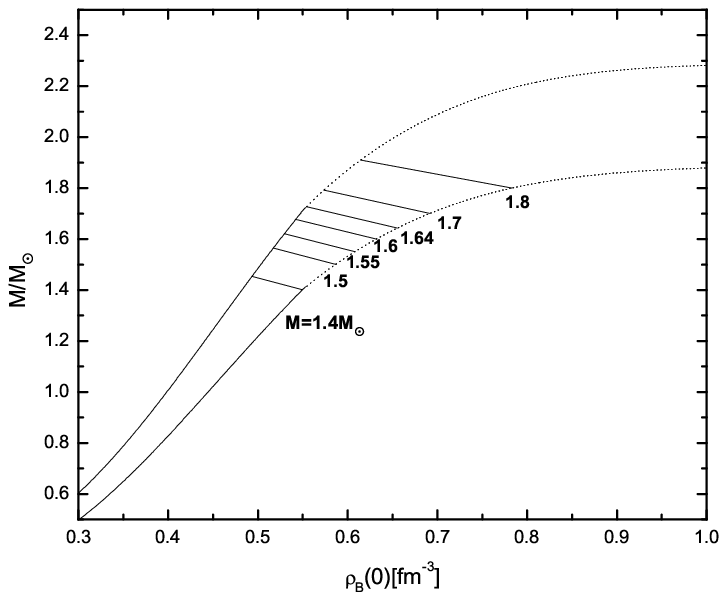}
   \caption{Gravitational mass M in solar masses as a function of the central
   density for rotating neutron star configurations with a deconfinement phase transition.
 The lower curve correspond to static configurations. the upper one to those with maximum rotation frequency $\nu_{k}$.
 The lines between both extremal cases connect configurations with the same total baryon number. The dotted lines indicate that the quark matter
 is produced in the core of neutron stars.The bag constant of the quark matter is B=108 $MeV fm^{-3}$}
   \label{Fig:f2}
   \end{figure}

\begin{figure}
   \centering
   \includegraphics[width=0.7\textwidth]{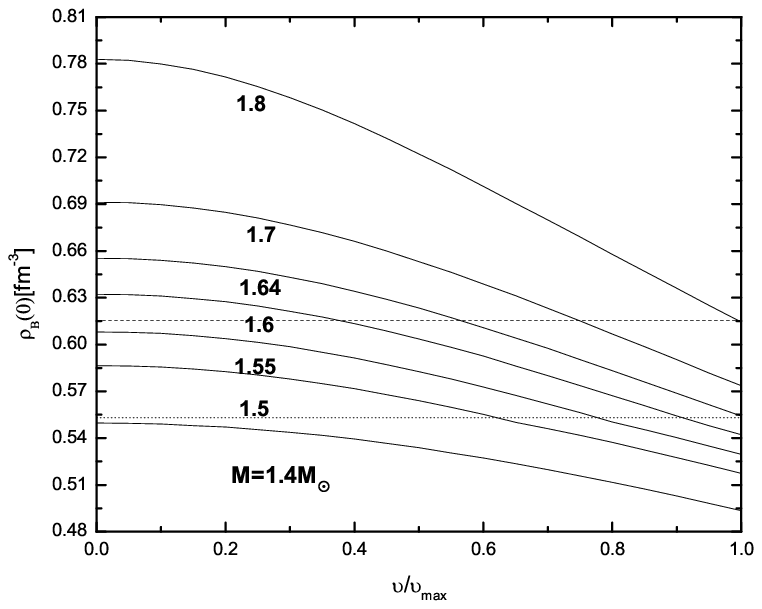}
   \caption{Central density as a function of rotational frequency for rotating
neutron stars of different gravitational mass at zero spin. All
sequences are with constant total baryon number. Dotted horizontal
lines indicate that deconfined quark matter is produced and dashed
horizontal lines indicate that the nucleon direct Urca process is
triggered.
 }
   \label{Fig:f3}
   \end{figure}

\begin{figure}
\centering
   \includegraphics[width=0.7\textwidth]{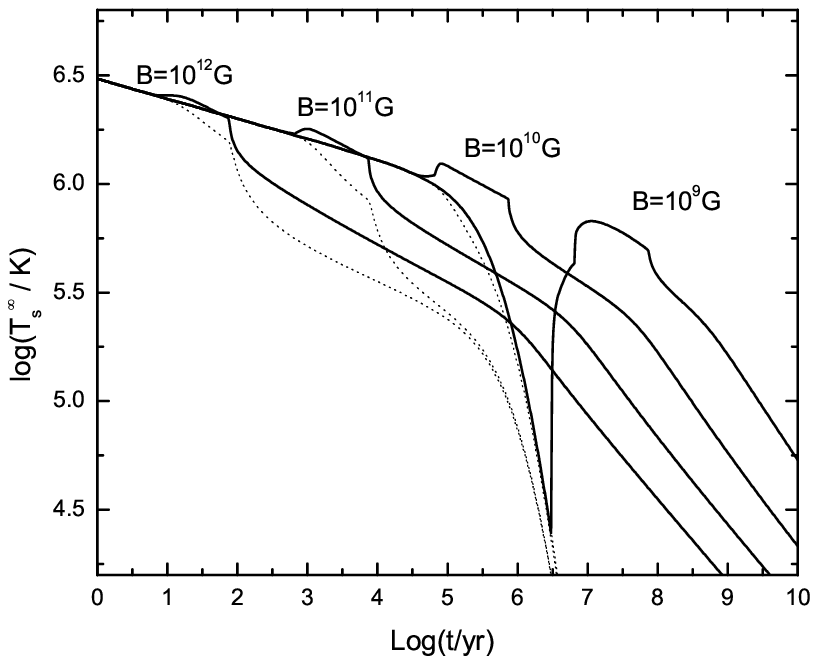}
   \caption{Thermal evolution curves of a 1.6 $M_{\odot}$ neutron star with
   deconfinement heating
   for various magnetic field strengths(solid curves) and the curves without deconfinement heating (dotted
   curves)}
   \label{Fig:f4}
   \end{figure}


\begin{figure}
   \centering
   \includegraphics[width=0.7\textwidth]{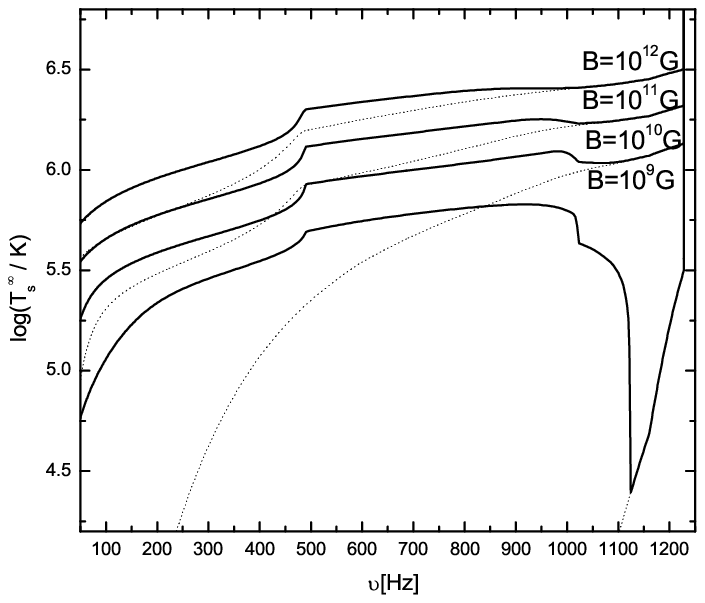}
   \caption{Surface temperature change of a 1.6 $M_{\odot}$ neutron star with rotational frequency  for
 various magnetic field strengths  with deconfinement heating
    (solid curves) and the curves without deconfinement heating (dotted
   curves)
 }
   \label{Fig:f5}
   \end{figure}

\begin{figure}
   \centering
   \includegraphics[width=1.0\textwidth]{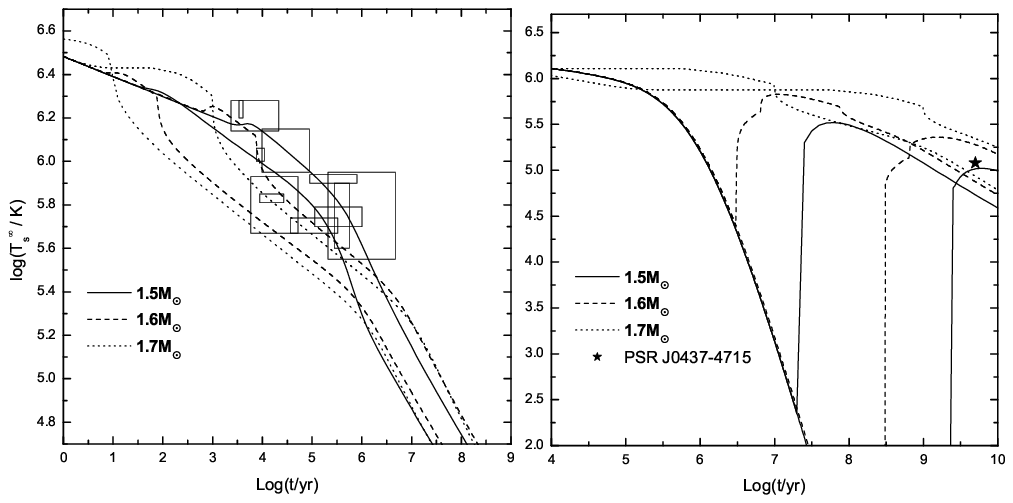}
   \caption{Thermal evolution curves of neutron stars with deconfinement heating for different stars masses and B=$10^{12}$,$10^{11}$G(left panel) and
   B=$10^{9}$,$10^{8}$G(right panel. Rectangles in the left-hand panel indicate observational data on cooling neutron stars with strong magnetic fields.)
 }
   \label{Fig:f6}
   \end{figure}
\end{document}